\documentclass[12pt]{article}
\usepackage{graphicx,amsmath,amssymb,url,enumerate,mathrsfs,epsfig,color}
\usepackage{tikz}
\usepackage{float}

\usepackage{amsmath}
\usepackage{amssymb}
\usepackage{yfonts}

\usepackage{tikz,pgfplots}

\usetikzlibrary{calc, patterns,arrows.meta, arrows, shapes.geometric}

\usepackage{graphicx,amsmath,amssymb,url,enumerate,mathrsfs,epsfig,color}

\usetikzlibrary{decorations.text}
\usetikzlibrary{decorations.markings}

\pgfplotsset{compat=1.8}

\usepackage{enumerate}
\usepackage{boxedminipage}
\usepackage{makeidx}
\usepackage{multicol}
\usepackage{xcolor}
\usepackage{mathtools}

\usetikzlibrary{calc, patterns,arrows, shapes.geometric}

\def\QED{\hskip0.1em\hfill\null\ \null\nobreak\hfill
\kern3pt\lower1.8pt\vbox{\hrule\hbox
{\vrule\kern1pt\vbox{\kern1.7pt \hbox{$\scriptstyle
QED$}\kern0.2pt}\kern1pt\vrule}\hrule}}

\newtheorem{rem}{Remark}[section]

\title{A buckling question}
\author{Marcelo Epstein\footnote{University of Calgary, Calgary, Canada. Email: mepstein@ucalgary.ca} }

\date{}

\begin{document}
\maketitle

\begin{abstract}
Although it is often asserted that, in view of their reduced length, axially compressible beams have a higher buckling load than their inextensible counterpart, a detailed analysis demnstrates that this is not necessarily the case. The argument to arrive at this conclusion is made in terms of relatively straightforward concepts of elasticity and structural mechanics. It is shown that for certain classes of materials the reduced pre-buckling length is more than compensated by a softening of the elastic response, leading to a reduction of the Euler critical load.
\end{abstract}


\section{Introduction}

The buckling load of an axially inextensible beam was determined by Euler in 1744, in an addendum to his book on isoperimetric problems of the calculus of variations. Euler, of course, does much more than finding this so-called critical load, but he does not consider the possible influence of the axial extensibility on the result. How then does the beam axial extensibility affect the buckling load? When asked this question, the usual answer of conscientious structural engineers is that the result is higher than Euler's critical load. The reason for this, it is argued, is that the compressive force prior to the onset of instability causes a reduction of the beam length. Since Euler's formula has the square of the beam length in the denominator, this must surely imply an increase in the magnitude of the critical load. 

This astute answer seems to imply that all one has to do to obtain the new critical load is to replace the original length with the new (reduced) length in the denominator of Euler's formula. That this is not exactly the case is demonstrated in many quantitative treatments, including \cite{magnusson, mazzilli, emam}.\footnote{Some of these treatments include the softening influence of shear strains.} Filipich and Rosales \cite{filipich} have also explicitly shown how the critical load depends on the stress-strain relation used. In particular, somewhat unexpectedly, it turns out that for some elastic materials the critical load is smaller in magnitude than the Euler load, thus proving that the intuitive reasoning described above is only partially correct. One may argue that for very small strains all constitutive laws with the same initial slope should give rise to the same results. Nevertheless, upon reflection, one arrives at the concusion that not only the slope but also the curvature of the stress-strain relation plays a role in the determination of the critical load. Indeed, the bending stresses brought about by the slight deviation from the straight configuration are superimposed on the previously applied compressive axial stress, whereby the slope of the stress-strain relation would have changed.

The purpose of this note is to derive the critical load for a compressible beam with a minimum of mathematical apparatus so as to be at the same time accurate and convincing. The stress will be regarded as a force per unit area, rather than as a twice contravariant component of a tensor in a not necessarily Cartesian coordinate system. The two contributors mentioned above, namely the pre-buckling shortening and the nature of the stress-strain relation, will be carefully implemented in Euler's formula to reproduce the results of \cite{filipich} in as simple and intuitive a manner as possible without resorting to the general continuum mechanics treatment.

\section{Stress-strain relations}
\label{sec:stress}

In the standard uniaxial material test, a cylindrical bar is subjected to a slowly varying axial force $F$. Measuring the length $L$ of the bar for each value of the applied force, a force-length graph is obtained for the material at hand. If the material is isotropic, the result on identical cylinders should be independent of the direction in which the cylinder was extracted from a presumptive three-dimensional block of material. The material is elastic if a removal of the load at any stage results in an immediate return of the sample to its original length. 

The passage from the uniaxial experiment to  a more general stress-strain relation is not completely straightforward, even in the uniaxial context. If the stress is the measured force divided by an area, which area? If the strain is the elongation divided by a length, which length? The Poisson effect results in a change of the cross-sectional area of the rod. In classical beam theory it is usually assumed that plane sections remain rigid, while preserving their perpendicularity to the deformed axis. The rigidity assumption implies the preservation of the area of the cross section, as if the Poisson effect were absent. The stress $\sigma$ in the putative uniaxial experiment is, therefore, uniquely defined as the centrally applied force divided by the area $A$ of the cross section.

The definition of uniaxial strain is somewhat subtler. Assuming the bar to be unstressed in its original configuration (before the application of the force), we can use the original length $L_0$ as reference and define the strain $\varepsilon$ as
\begin{equation} \label{eq1}
\varepsilon=\frac{L-L_0}{L_0}=\frac{e}{L_0},
\end{equation}
where $e$ represents the elongation measured from the unstressed configuration. Many other definitions are possible and valid. We remark that since the length of the deformed rod must always remain positive, the physically meaningful range of $\varepsilon$ is $(-1,\infty)$,

In many applications, it is also useful to define a strain relative to a different reference length, $\hat L$ say, which is not stress free. The elongation relative to $\hat L$ is given by
\begin{equation}
{\hat e} = e -({\hat L}-L_0) =e-\varepsilon_0 L_0,
\end{equation}
where $\varepsilon_0$ is the strain of the configuration of length $\hat L$, according to Equation (\ref{eq1}).
The corresponding measure of strain relative to $\hat L$ is
\begin{equation} \label{eq3}
{\hat \varepsilon}=\frac{{\hat e}}{\hat L}= \frac{\varepsilon - \varepsilon_0}{1+\varepsilon_0}.
\end{equation}

An elastic constitutive equation in dimension one boils down to a single function that expresses the stress as a function of a strain measure, such as
\begin{equation}
\sigma=f(\varepsilon).
\end{equation}
The {\it tangent or instantaneus modulus of elasticity} $E_p$ at the strain $\varepsilon_p$ is the derivative of the stress-strain relation at that point, namely,
\begin{equation}
E_p=\left. \frac{df(\varepsilon)}{d\varepsilon}\right|_{\varepsilon=\varepsilon_p}.
\end{equation}
The tangent modulus of elasticity at $\varepsilon=0$ is denoted by $E_0$.

Analogously, if the stress-strain relation is given as
\begin{equation}
\sigma=g({\hat\varepsilon}),
\end{equation}
that is,  in terms of the strain $\hat\varepsilon$ measured relative to the length $\hat L$, the corresponding tangent modulus at a strain ${\hat\varepsilon}_q$ is defined as
\begin{equation}
{\hat E}_q=\left. \frac{dg({\hat\varepsilon})}{d{\hat\varepsilon}}\right|_{{\hat\varepsilon}={\hat\varepsilon}_q}.
\end{equation}
If $q$ corresponds to $p$, namely if, in accordance with Equation (\ref{eq3}),
\begin{equation} \label{eq3}
{\hat \varepsilon}_q= \frac{\varepsilon_p - \varepsilon_0}{1+\varepsilon_0},
\end{equation}
we obtain
\begin{equation}
{\hat E}_q=(1+\varepsilon_0) E_p.
\end{equation}
In other words, the two stress-strain graphs are not merely related by a translation, but also by a re-scaling of the abscissas.

\section{The extended Euler formula}

With the meagre theoretical background above described, we can generalize Euler's celebrated formula for the critical load of an inextensible pinned elastic bar of constant cross section, viz.
\begin{equation} \label{eq10}
P_{cr} = - \frac{\pi^2 E_0  I}{L_0^2},
\end{equation}
where the negative sign to indicate compression is added for convenience. In Euler's formula (\ref{eq10}), $I$ denotes the smaller principal moment of inertia of the cross section. 

We recall that Euler's formula can be obtained by perturbing the straight configuration by means of a small transverse deflection function, and considering the bending moment distribution induced upon this perturbation by the applied axial load. To find a non-trivial equilibrium configuration of this kind a classical eigenvalue problem must be solved for the governing 4th order differential equation with the corresponding homogeneous boundary conditions. Applying the same reasoning to a compressed elastic bar, even if the axial strain $\varepsilon_{cr}$ under the applied force $P_{cr}$ is of very large magnitude, while the transverse perturbation is infinitesimal, we obtain a similar eigenvalue problem except for the following two details:
\begin{enumerate}
\item The boundary conditions of support apply to the ends of a bar with the contracted length ${\hat L}=L_0(1+\varepsilon_{cr})$,
\item The small increments of stress due to bending are governed by the tangent modulus $\hat E$ evaluated at the contracted configuration.
\end{enumerate}
Thus, the extended Euler formula can be written as
\begin{equation}
{\hat P}_{cr}=- \frac{\pi^2 {\hat E} I}{{\hat L}^2}.
\end{equation}
On the basis of Section \ref{sec:stress}, we can write
\begin{equation} \label{eq12}
{\hat P}_{cr}=-\frac{\pi^2E_{cr} I}{L_0^2 (1+\varepsilon_{cr})}=-\frac{\pi^2\left(\frac{d\sigma}{d\varepsilon}\right)_{cr} I}{L_0^2 (1+\varepsilon_{cr})}
\end{equation}

\section{Further considerations and examples}

In Equation (\ref{eq12}) we can identify two elements that affect the critical load when compared with the classical Euler counterpart. The denominator contains the factor $(1+\varepsilon_{cr})$. Since $\varepsilon_{cr}$ is negative (with an absolute value smaller than $1$), the influence of this factor is always an increase of the absolute value of the critical load. 

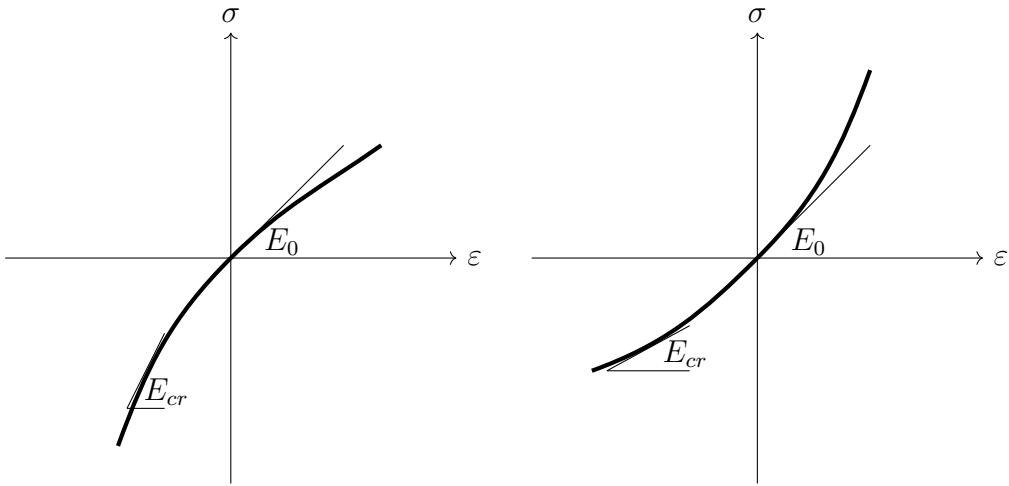
\begin{figure}[H]
\begin{center}
\begin{tikzpicture} [scale=1.0]
\draw[->] (0,0)--(6,0);
\node[right]  at (6,0) {$\varepsilon$};
\draw[->] (3,-3)--(3,3);
\node[above] at (3,3) {$\sigma$};

\draw[ultra thick] (1.5,-2.5) to [out=70, in=225] (3,0);
\draw[ultra thick] (3,0) to [out=45, in=215] (5,1.5);
\draw (1.62,-2)--(2.12,-1);
\draw (1.62,-2)--(2.12,-2);
\node[above right] at (1.7,-2.1) {$E_{cr}$};
\draw (3,0)--(4.5,1.5);
\node [above right] at (3.3,-0.1) {$E_0$};

\begin{scope} [xshift=7cm]
\draw[->] (0,0)--(6,0);
\node[right]  at (6,0) {$\varepsilon$};
\draw[->] (3,-3)--(3,3);
\node[above] at (3,3) {$\sigma$};

\draw[ultra thick] (0.8,.-1.5) to [out=20, in=225] (3,0);
\draw[ultra thick] (3,0) to [out=45, in=250] (4.5,2.5);
\draw (1.0,-1.5)--(2.1,-0.9);
\draw (1.0,-1.5)--(2.1,-1.5);
\node[below right] at (1.60,-0.95) {$E_{cr}$};
\draw (3,0)--(4.5,1.5);
\node [above right] at (3.3,-0.1) {$E_0$};
\end{scope}

\end{tikzpicture}
\end{center}
\caption{Compressive stiffening (left) and softening (right) responses}
\label{fig1}
\end{figure}

The second element contributing to the change of the final result is the presence in the numerator of (\ref{eq12}) of the tangent modulus of elasticity at the compressed configuration of the bar. Depending on the properties of the stress-strain relation being used, this second element may contribute to a decrease of the critical load or to its further increase. Figure \ref{fig1} illustrates two possible behaviours of a material in uniaxial compression, namely, stiffening and softening responses.

To explore the possible consequences of the curvature of the stress-strain relation upon the onset of elastic instability, one may consider a quadratic stress-strain relation of the form
\begin{equation} \sigma=E_0 (\varepsilon + \alpha \varepsilon^2),
\end{equation}
where $\alpha$ is a unitless coefficient.
If $\alpha > 0$ we obtain softening compressive behaviour.  The tangent modulus for this class of materials is given by $E=E_0 (1+2 \alpha \varepsilon)$.  The range of applicability of this constitutive law (with $\alpha>0$) is the interval $-1/(2\alpha) < \varepsilon < \infty$.

\begin{rem}{\rm {\bf A balanced material}: The particular case $\alpha=0.5$, when plugged into Equation (\ref{eq12}), delivers the somewhat surprising result that, for this special choice of stress-strain relation, the Euler critical load is recovered. The effect of length reduction is thus exactly compensated by the compressive softening. For $\alpha > 0.5$, therefore, the overall combined effect of the reduced length and the compressive softening results in a net reduction of the absolute value of the critical load. We may refer to these materials as {\it sub-Eulerian}.}
\end{rem}

Given a specific stres-strain relation $\sigma=f(\varepsilon)$, Equation (\ref{eq12}) implies that the strain $\varepsilon_{cr}$ at the critical load of the compressible bar is obtained as the smallest (in absolute value) negative real root of the algebraic equation
\begin{equation} \label{eq14}
(1+\varepsilon) f(\varepsilon) + \pi^2 s^2 f'(\varepsilon) = 0,
\end{equation}
where primes denotes derivatives with respect to $\varepsilon$, and where $s =\sqrt{ I /AL_0^2}$ is the reciprocal of the slenderness ratio of the bar. We remark that the product $\varepsilon_e=-\pi^2s^2$ can be interpreted as the strain associated to the Euler inextensible solution. For purposes of comparison, the critical load ${\hat P}_{cr}$ can be conveniently expressed as a multiple of $E_0 I/L_0^2$. In this scale, the classical Euler load is represented by $-\pi^2=-9.8696$.

\subsection{Hooke's law}

Under the standard Hookean constitutive equation
\begin{equation}
\sigma=E\varepsilon,
\end{equation}
where $E$ is constant, Equation (\ref{eq14}) becomes
\begin{equation}
(1+\varepsilon) \varepsilon + \pi^2 s^2 = 0.
\end{equation}
The physically meaningful root of this quadratic equation is
\begin{equation} \label{eq17}
\varepsilon_{cr} = - \frac{1-\sqrt{1-4 \pi^2 s^2 }}{2},
\end{equation}
a well-known formula \cite{mazzilli}. The slenderness ratio must be not smaller that $2\pi$ for the rod to buckle at all.

In \cite{filipich} a slenderness parameter $q$ is defined as $q^2=12 s^2$. For $q=0.35$ we obtain, according to Equation (\ref{eq17}), $\varepsilon_{cr}=-0.113674$. Finally,
\begin{equation}
{\hat P}_{cr}=E_0A \varepsilon_{cr}=- 11.1354\, \frac{ E_0 I}{L_0^2}.
\end{equation}

\subsection{ St. Venant-Kirchhoff law}

A frequently used constitutive law in non-linear elasticity establishes a linear relation between the second Piola-Kirchhoff stress tensor and the non-linear Lagrange strain tensor. Regardless of the terminology employed and of the viability of its use for very large compressive strains, it constitutes a practical option for materials that exhibit softening compressive behaviour provided that the strains are of moderate magnitude (say up to $20\%$). Be that as it may, in the context of our present interest, the stress-strain relation is expressed as the cubic polynomial
\begin{equation} \label{eq19}
\sigma=f(\varepsilon)=E_0  (\varepsilon+1)(\varepsilon+0.5 \varepsilon^2).
\end{equation}
Introducing this expression into Equation (\ref{eq14}), we obtain the algebraic equation
\begin{equation} \label{eq20}
(1+\varepsilon)^2 (\varepsilon+0.5 \varepsilon^2) + \pi^2 s^2 (1+3 \varepsilon+1.5 \varepsilon^2) = 0,
\end{equation}
For $q=0.35$, as in our previous example, the desired root of this equation is $\varepsilon_{cr}=-0.0941677$. The corresponding critical load is obtained from Equation (\ref{eq19}) as ${\hat P}_{cr}=- 0.0812839 \, E_0 A$, which can be written as
\begin{equation}
{\hat P}_{cr}=E_0A \varepsilon_{cr}=- 7.9625\, \frac{ E_0 I}{L_0^2}.
\end{equation}
Due to the softening behaviour of this material in compression, the critical load of the extensible bar, in spite of its diminished pre-buckling length, is reduced to about $80\%$ of the Euler load.

\subsection{A logarithmic law}

The Hencky logarithmic strain measure $\ln(1+\varepsilon)$ has many desirable features, such as being potentially applicable over the whole physical range $-1 < \varepsilon < \infty$. It is, therefore in widespread use in many applications. The simplest stress-strain relation in terms of this strain measure is
\begin{equation}
\sigma=E_0 \ln(1+\varepsilon).
\end{equation}
From Equation (\ref{eq14}), the critical strain is the solution of the equation
\begin{equation} \label{eq23}
(1+\varepsilon) \ln(1+\varepsilon) + \pi^2 s^2/(1+\varepsilon) = 0.
\end{equation}
Proceeding as before, with $q=0.35$, we obtain $\varepsilon_{cr}=-0.122699$, which leads to the final result
\begin{equation}
{\hat P}_{cr}=E_0A\ln(1+\varepsilon_{cr})=- 12.8234\, \frac{ E_0 I}{L_0^2}.
\end{equation}

\section{Summary and conclusions}

The proposed interpretation of a modified version of the classical Euler load at the onset of elastic instability of a straight elastic extensible beam incorporates the two aspects that contribute to an increase or decrease of this load. The first factor, usually the only one adduced intuitively in support of the increase of the critical load, is that the shortened length of the beam prior to the onset of instability surely implies a reduction of the beam slenderness, rendering it more stable. The second factor is of a more subtle nature as it involves, even within the realm of small strains and certainly in the case of large ones, a possible softening of the instantaneous modulus of elasticity of the material. This second factor may be large enough to counteract the reduction in length and result in a reduced intensity of the critical load.

One of the features of the approach presented in this note is that it can be derived and explained in relatively simple engineering terms, as opposed to derivations based on the more refined and involved concepts of non-linear continuum mechanics. Studies of this last type, such as the excellent work of Filipich and Rosales \cite{filipich}, provide also a description of many other important phenomena associated with instability, including the intricate post-buckling behaviour of the bar. Nevertheless, as far as the critical loads are concerned, the results produced by the relatively simple approach introduced in this note are identical to those of \cite{filipich} for all the materials considered therein.

\end{document}